\documentclass[12pt]{article}
\usepackage{epsfig}
\usepackage{pst-plot,url,epsf}
\usepackage{t1enc}
\newcommand{\beq}{\begin{equation}}
\newcommand{\eeq}{\end{equation}}
\newcommand{\bea}{\begin{eqnarray}}
\newcommand{\eea}{\end{eqnarray}}

\newcommand{\epm}{e^+e^-}
\newcommand{\nn}{\nonumber}

\newcommand{\ra}{\rightarrow}
\def\earr{\end{array}}
\def\barr#1{\begin{array}{#1}}

\begin{document}
\thispagestyle{empty}
\begin{flushright}
March 2013\\
Revised version:\\
June 2013\\
\vspace*{1.5cm}
\end{flushright}
\begin{center}
{\LARGE\bf Secondary lepton distributions as a probe of the top--higgs 
           coupling at the LHC}\\
\vspace*{2cm}
K. Ko\l odziej\footnote{E-mail: karol.kolodziej@us.edu.pl}\\[1cm]
{\small\it
Institute of Physics, University of Silesia\\ 
ul. Uniwersytecka 4, PL-40007 Katowice, Poland}\\
\vspace*{4.5cm}
{\bf Abstract}\\
\end{center}
The differential distributions in rapidity and angles of the secondary lepton 
in the associated production of the top quark pair and higgs boson 
in proton--proton collisions at the LHC are quite sensitive to the top--higgs
coupling. However, the effects of anomalous couplings of the most general 
$t\bar t h$ interaction with operators of dimension-six that are clearly 
visible in the signal of the associated production of the top quark pair and 
higgs boson are to large extent obscured by the background
sub-processes with the same final state. This means that analyses of such 
effects, in addition to higher order corrections that are usually 
calculated for the on-shell top quarks and higgs boson, 
should include their decays and possibly complete off resonance background 
contributions to the corresponding exclusive reactions.

\vfill

\newpage

\section{Introduction.}

Associated production of the top quark pair and higgs boson was proposed
as a sensitive probe of the top--higgs Yukawa coupling $g_{tth}$ at the
$\epm$ linear collider (LC) \cite{ILC}, \cite{CLIC} more than 20 years ago 
\cite{gtth}. A clean experimental environment of the LC seems
to be the best place to study the higgs boson profile, including the
measurement of $g_{tth}$, but the project of LC is still at the rather early 
stage of TDR. Fortunately, the top quarks are copiously produced at the LHC
that, among others, allows for more and more precise determination 
of the top quark pair production cross section and for measurements of the
cross sections  
of $t\bar t\;+$~jets, see \cite{schilling} for a review. The 
measurement of production of $t\bar t+b\bar b$ \cite{CMS} is
particularly interesting, as it is relevant for observation of the associated 
production of the top quark pair and higgs boson, with the higgs decaying
into $b\bar b$. The latter should be the dominant decay mode, if the new
boson at a mass of about 125~GeV observed at the LHC \cite{higgs}
is indeed the higgs.

\begin{figure}[htb]
\centerline{
\epsfig{file=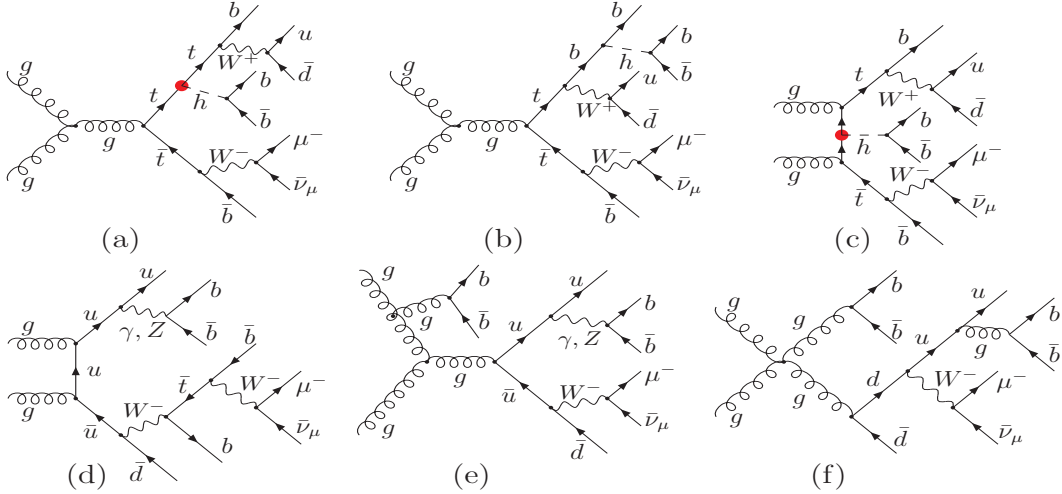,  width=140mm, height=65mm}}
\caption{Examples of the lowest order Feynman diagrams of reaction 
(\ref{ggbbbudbmn}): (a), (b) and (c) are the signal diagrams of $t\bar t h$ 
production, (d), (e) and (f) are the $t\bar t h$ background contributions. 
Blobs indicate the higgs--top coupling.}
\label{diags}
\end{figure}

The associated production of the top quark pair and higgs boson in the
proton--proton collisions at the LHC 
\bea
\label{pphtt}
pp \;\ra\; t \bar t h 
\eea
is dominated by the gluon--gluon fusion mechanism. Taking into
account decays: $h\to b\bar b$, $t\to bW^+$, $\bar t\to \bar bW^+$
and the subsequent decays of the $W$-bosons, one should consider hard
scattering partonic processes as, e.g.,
\bea
\label{ggbbbudbmn}
gg \;\ra\; b u \bar{d} \bar b \mu^- \bar \nu_{\mu} b \bar b ,
\eea
corresponding to one of the $W$'s decaying hadronically and the other 
leptonically. Reaction (\ref{ggbbbudbmn}) receives contributions from
$67\,300$ Feynman diagrams in the leading order of the standard model (SM),
in the unitary gauge neglecting masses smaller than $m_b$, of
which barely 56 diagrams constitute the signal of the $t \bar th$ production 
and subsequent decay. Some examples of the Feynman diagrams of 
(\ref{ggbbbudbmn}) are shown in Fig.~\ref{diags}. The 56 signal diagrams
are obtained from those depicted in Fig.~\ref{diags}(a), \ref{diags}(b)
and \ref{diags}(c) by attaching the $hb\bar b$-vertex to other top
or bottom quark lines, interchanging external $b$ and $\bar b$ quarks 
in each of the figures and interchanging the
two initial state gluons of Fig.~\ref{diags}(c). The diagrams shown
in Fig.~\ref{diags}(d), \ref{diags}(e) and \ref{diags}(f) are just a few
examples of the background contributions to associated production
of the higgs boson and top quark pair.

A question arises whether the associated
production of the higgs boson and top quark pair can be sensitive to
possible modifications of the SM top--higgs Yukawa coupling or not.
The question will be addressed in this work by showing how the
distributions of the secondary lepton are changed in the presence
of such modifications. The distributions computed with the signal
diagrams only will be compared with those computed with the full set
of the leading order Feynman diagrams that will demonstrate how the
background contributions obscure relatively clear effects of the
anomalous $t\bar t h$ coupling in the signal cross section. 
Although the issue may seem somewhat premature from 
the experimental side, but in view of the excellent performance of the LHC, 
it may become relevant in quite a near future.

\section{Calculation details}

The calculation is performed in a fully automatic way with a new version 
\cite{carlomat2} of {\tt carlomat} \cite{carlomat}, a general purpose program 
for Monte Carlo computation of lowest order cross sections. 
The most general Lagrangian of $t\bar t h$ interaction including corrections
from dimension-six operators that has been implemented in the program has 
the following form \cite{aguilar}
\bea
\label{httcoupl}
\mathcal{L}_{t\bar t h}=-g_{t\bar th}\bar{t}\left(f+if'\gamma_5\right)t h,
\eea
where $g_{t\bar th}=m_t/v$, with $v=(\sqrt{2}G_F)^{-1/2}\simeq 246$~GeV, 
is the top--higgs Yukawa coupling. The couplings $f$ and $f'$ 
are assumed to be real. They describe, 
respectively, scalar and pseudoscalar departures from the purely scalar
top--higgs Yukawa coupling of SM. The latter is reproduced for $f=1$
and $f'=0$. Other dimension-six gauge-invariant effective operators that
may have affected the $t\bar t h$ interaction are redundant, in a sense that 
they can be eliminated with the use of the equations of motion, both for
the on- and off-shell particles \cite{aguilar}.
Obviously, the process of associated production of the higgs boson
and top quark pair will be affected by many other possible deviations
from the SM couplings. 
They are not considered here, as the primary goal of the present
work is to illustrate just the effects of the anomalous $t\bar t h$ interaction
on the distributions of the secondary lepton. However, some deviations, eg., 
the anomalous $Wtb$ coupling generated by the gauge-invariant  
dimension-six effective operators, which is present in the $t\bar t h$ signal
diagrams of Figs.~\ref{diags}(a)--(c) and in some off resonance background
diagrams such as the one depicted in Fig.~\ref{diags}(d),
can be easily included, as it
has been already implemented in {\tt carlomat}. See \cite{wtblhc} for 
the illustration of some effects on the top quark pair production at the 
LHC that can be caused by the $Wtb$ coupling.

The couplings $f$ and $f'$ of Lagrangian (\ref{httcoupl}) belong to least 
constraint couplings of the SM. For the higgs boson with a mass of 125~GeV, 
practically the only model independent way to constrain them is
to measure the $t\bar t h$ production \cite{dirconstr}. First results
of search for this process in $pp$ collisions at the
LHC are reported in \cite{tthCMS}.
Indirect constraints of the $t\bar t h$ interaction vertex can be derived 
from measurements of the higgs boson 
production rate through the gluon--gluon fusion process, which is dominated by
a top-quark loop, and of the higgs boson decay into 2 photons that,
despite being dominated by the $W$ boson loop, also receives 
a significant contribution from the top-quark loop. However, extraction 
of the $t\bar t h$
coupling in this way relies on the assumption that the loops do not
receive contributions from new massive fundamental particles beyond those of 
the SM. If two universal scale factors are assumed, one for the higgs boson
Yukawa couplings to all the SM fermion species and the other for
the higgs boson couplings  to the EW gauge bosons, and
if there is no new physical degrees of freedom, then the scalar coupling $f$
of Eq.~(\ref{httcoupl}) can be constraint at 95\% C.L. to be in the following
regions:
\bea
\label{fATLAS}
f\in [-1.2,-0.6]\cup[0.6,1.3]&&\qquad 
{\rm ATLAS}\;\cite{fATLAS}\\
\label{fCMS}
f\in [0.3,1.0]\qquad\qquad\qquad\; &&\qquad 
{\rm CMS}\;\cite{fCMS}.
\eea
It should be noted at this point that an opposite sign of the higgs boson 
coupling to fermions with respect to its coupling to the gauge bosons 
is required in the Lagrangian for the unitarity and renormalizability 
of the theory \cite{unitarity} and vacuum stability \cite{stability}.
Therefore, the interval in the range of negative numbers in (\ref{fATLAS})
is highly disfavoured. The relative sign of both couplings could probably be 
best determined in the reaction of associated production of 
the top quark and higgs boson in proton-proton collisions at the LHC through 
the underlying $t$-channel partonic process $qb\to tq'h$
\cite{biswas}.

In {\tt carlomat}, the on-shell poles in propagators of unstable particles,
both the $s$- and $t$-channel ones, are avoided by making the following 
substitutions:
\beq
\label{m2}
m_b^2 \;\ra \; M_b^2=m_b^2-im_b\Gamma_b, \quad b=Z, W, h,\qquad
m_t\;\ra \; M_t=\sqrt{m_t^2-im_t\Gamma_t},
\eeq
where the particle widths are assumed to be constant and the square root with
positive real part is chosen, see \cite{carlomat} for details. 
In order to minimize unitarity violation effects at high energies
caused by substitutions (\ref{m2}), which correspond to re-summation of one
particle irreducible higher order contributions to $s$-channel propagators,
the computation is performed in the complex mass scheme, where the electroweak 
(EW) couplings are parametrized in terms
the complex EW mixing parameter $\sin^2\theta_W=1-{M_W^2}/{M_Z^2}$
which preserves the lowest order Ward identities \cite{Racoon}. Note, that 
the electric charge
$e_W$ can be defined as a real quantity in terms of $\alpha_W$
\bea
\label{alphaw}
e_W=\sqrt{4\pi\alpha_W}, \qquad {\rm with} \qquad
\alpha_W=\frac{\sqrt{2}G_{F}m_W^2}{\pi}\left(1-\frac{m_W^2}{m_Z^2}\right),
\eea
as it enters all the EW couplings multiplicatively, 
which is our choice in the present work. The only effect of using the complex 
masses of (\ref{m2}) in Eq.~(\ref{alphaw}) would be the overall change of
normalization of the cross section.
The top--higgs Yukawa coupling is defined in the complex mass scheme by
\bea
\label{yuk}
      g_{t\bar th}=e_W\;\frac{M_t}{2\sin\theta_W M_W},
\eea
i.e., it is a complex quantity, as it is parametrized in terms of the
complex masses of (\ref{m2}) and complex EW mixing parameter $\sin\theta_W$.

\section{Results}

In this section some results for the differential cross sections and
distributions of reaction
\bea
\label{ppbbbudbmn}
pp \;\ra\; b \bar b b u \bar{d} \bar b \mu^- \bar \nu_{\mu}
\eea
at $\sqrt{s}=14$~TeV are presented. For the sake of simplicity and easy
reproducibility of the results, only one hard scattering process 
(\ref{ggbbbudbmn}) that dominates at that energy
is taken into account. It is folded with CTEQ6L parton distribution 
functions \cite{CTEQ} at the scale $Q = 2m_t+m_h$.

The initial physical input parameters used in the computation are the 
following. 
The strong coupling between quarks and gluons is given by 
$g_s=\sqrt{4\pi\alpha_s}$, with $\alpha_s(m_Z)=0.118$. The EW couplings
are parametrized in terms of the electric charge of (\ref{alphaw}) that
is kept real and the complex EW mixing parameter $\sin\theta_W$,
as described in Section 2, with the EW gauge boson masses and widths: 
$m_W=80.419$~GeV, $\Gamma_W=2.12$~GeV, $m_Z=91.1882$~GeV, $\Gamma_Z=2.4952$~GeV
and the Fermi coupling $G_F=1.16639 \times 10^{-5}\;{\rm GeV}^{-2}$.
The top quark and higgs boson masses are: $m_t=173$~GeV, $m_h=125$~GeV
and their widths that are calculated to the lowest order of SM are the 
following: $\Gamma_t=1.49165$~GeV, $\Gamma_h=4.9657$~MeV.
The $b$-quark and muon masses are also kept non zero, but their
actual values: $m_b=4.5\;{\rm GeV}$ and $m_{\mu}=105.65837\;{\rm MeV}$, are
numerically irrelevant in practise. Masses of the light quarks of 
(\ref{ppbbbudbmn}) are neglected.

Jets are identified with their original partons and the following 
cuts on the transverse momenta $p_T$, pseudorapidities $\eta$, 
missing transverse energy $/\!\!\!\!E^T$ and separation $\Delta R_{ik}$ 
in the pseudorapidity--azimuthal angle $(\varphi)$ plane between 
the objects $i$ and $k$  are imposed:
$$p_{Tl} > 30\;{\rm GeV}, \qquad p_{Tj} > 30\;{\rm GeV}, \qquad
\left|\eta_l\right| < 2.1, \qquad \left|\eta_j\right| < 2.4,$$
\bea
\label{cuts}
/\!\!\!\!E^T > 20\;{\rm GeV},\qquad
\Delta R_{lj,jj}=
\sqrt{\left(\eta_i-\eta_k\right)^2+\left(\varphi_i-\varphi_k\right)^2} \;>\; 0.4,
\eea
where the subscripts $l$ and $j$ stand for {\em lepton} 
and {\em jet}. Cuts (\ref{cuts}) should allow to select events with separate 
jets, an isolated charged lepton and missing transverse momentum.

Moreover, 100\% efficiency of $b$ tagging is assumed and  events of the 
associated production of top quark pair and higgs boson in
reaction (\ref{ppbbbudbmn}) are selected by imposing the following invariant 
mass cuts:
on the invariant mass of two non $b$ jets, $b_{\sim b_1}$ and $b_{\sim b_2}$,
\beq
\label{cutmw}
60\;{\rm GeV} < \left[\left(p_{\sim b_1}+p_{\sim b_2}\right)^2\right]^{1/2} < 90
\;{\rm GeV},
\eeq
on the transverse mass of the muon--neutrino system
\bea
\label{cutmwt}
\left[m_{\mu}^2+2\left(m_{\mu}^2+\left|{\vec p}_{\mu}^{\;T}\right|^2\right)^{1/2}
\left|\;/\!\!\!{\vec p}^{\;T}\right|
-2{\vec p}_{\mu}^{\;T}\cdot /\!\!\!{\vec p}^{\;T}\right]^{1/2} < 90\;{\rm GeV},
\eea
on the invariant mass of a $b$ jet, $b_1$, and the two non 
$b$ jets
\beq
\label{cutmt}
\left|\left[\left(p_{b_1}+p_{\sim b_1}+p_{\sim b_2}\right)^2\right]^{1/2} - m_t\right|
< 30\;{\rm GeV},
\eeq
on the transverse mass $m_T$ of a $b$ quark, $b_2$, 
muon and missing transverse energy
\beq
\label{cutmtt}
m_t-30\;{\rm GeV} < m_T < m_t+10\;{\rm GeV}
\eeq
and the invariant mass cut on two $b$ jets, $b_3$ and $b_4$, 
\beq
\label{cutmh}
\left|\left[\left(p_{b_3}+p_{b_4}\right)^2\right]^{1/2} - m_h\right|
< m_{bb}^{\rm cut}, 
\eeq
with either $m_{bb}^{\rm cut} = 20$~GeV or, more optimistically, 
$m_{bb}^{\rm cut}=10$~GeV.
In (\ref{cutmtt}), $m_T$ is the transverse mass defined by
\bea
m_T^2=m^2+2\left(m^2+\left|{\vec p}_{b_2}^{\;T}+{\vec p}_{\mu}^{\;T}
\right|^2\right)^{1/2}/\!\!\!\!E^T
-2\left({\vec p}_{b_2}^{\;T}+{\vec p}_{\mu}^{\;T}\right)
\cdot /\!\!\!{\vec p}^{\;T}, \nn
\eea
with $m$ being the invariant mass of the $b$-$\mu$ system given by
$m^2=\left(p_{b_2}+p_{\mu}\right)^2$. Cuts (\ref{cutmw})--(\ref{cutmtt})
should allow to identify the secondary $W$ bosons, the top quarks and
the higgs boson. They were used before in the context of the associated 
production of the top quark pair and higgs boson in $\epm$ collisions 
at the LC \cite{KS}.

For the sake of illustration, the $t\bar th$ couplings of (\ref{httcoupl}) 
are assigned the following values: $f=1,0$ and $f'=0,\pm 1$ and 
the differential distributions of the final state
muon, generally referred to as {\em lepton}, of reaction (\ref{ppbbbudbmn})
are computed, first with the 56 signal Feynman diagrams of the 
associated production of the top quark pair and higgs boson 
and then with the
complete set of $67\,300$ Feynman diagrams, as discussed in Section 1.
The rapidity and angular differential cross sections and distributions of 
the lepton for which the effects of anomalous couplings are best visible  
will be shown in Figs.~\ref{figrapl}--\ref{fig2d}
and the distributions in the lepton transverse momentum 
or energy which are practically not affected by the couplings will not 
be presented.

\begin{figure}[htb]
\vspace*{0.5cm}
\begin{center}
\setlength{\unitlength}{1mm}
\begin{picture}(35,35)(0,0)
\includegraphics{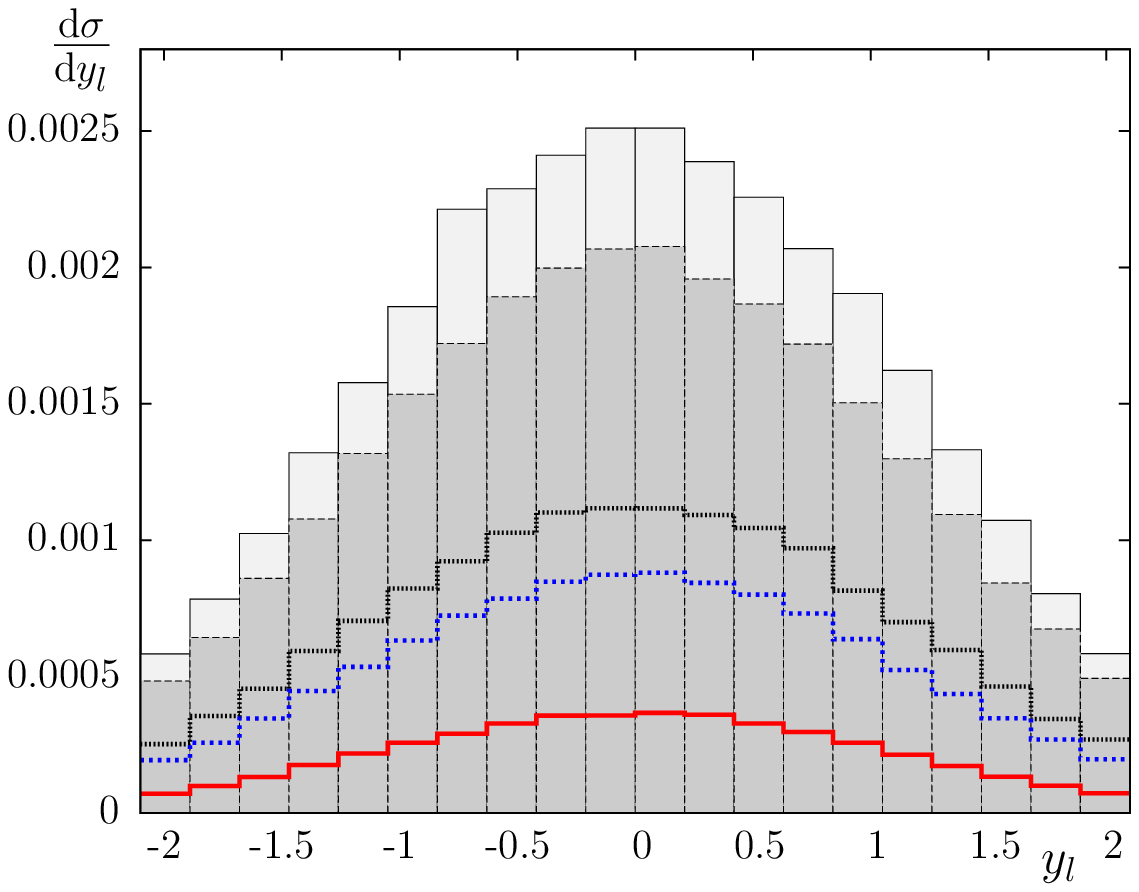}
\end{picture}
\hspace*{2.cm}
\begin{picture}(35,35)(0,0)
\includegraphics{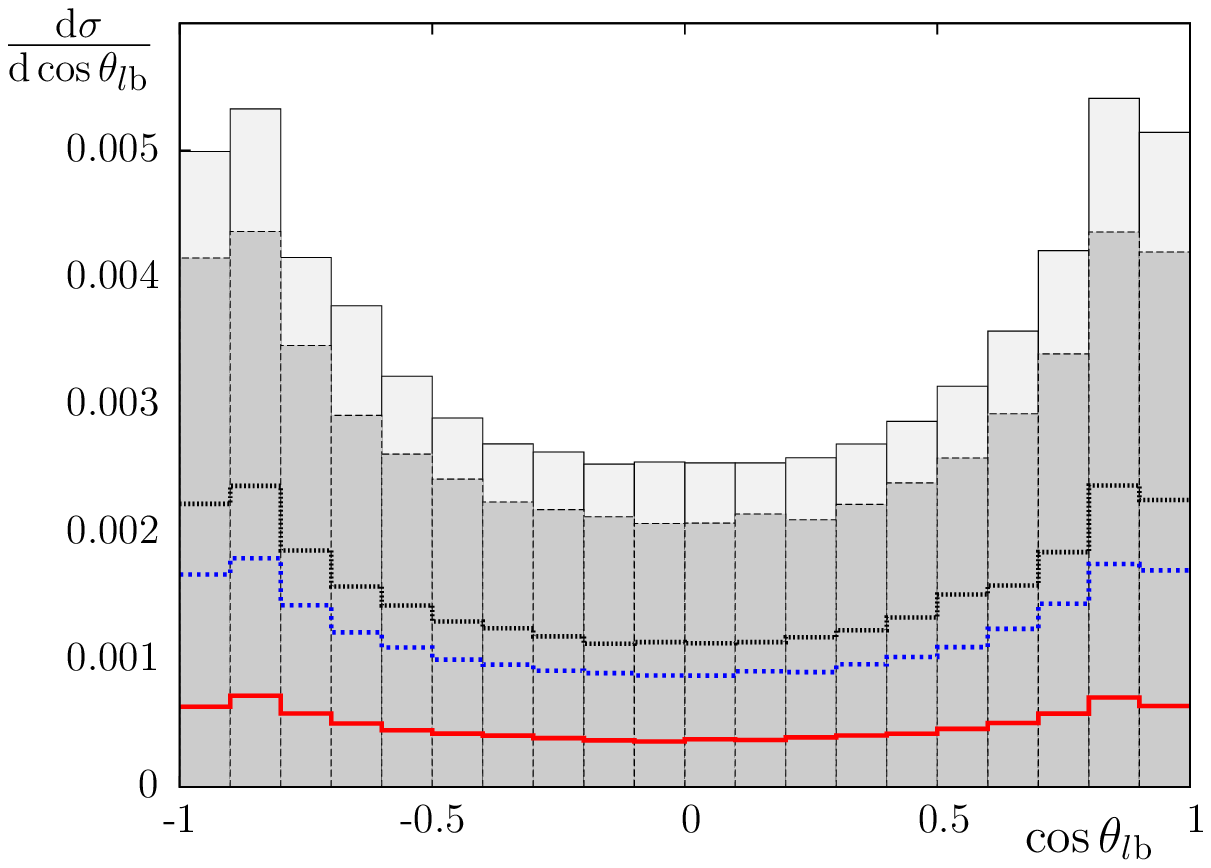}
\end{picture}
\hfill
\begin{picture}(35,35)(0,0)
\includegraphics{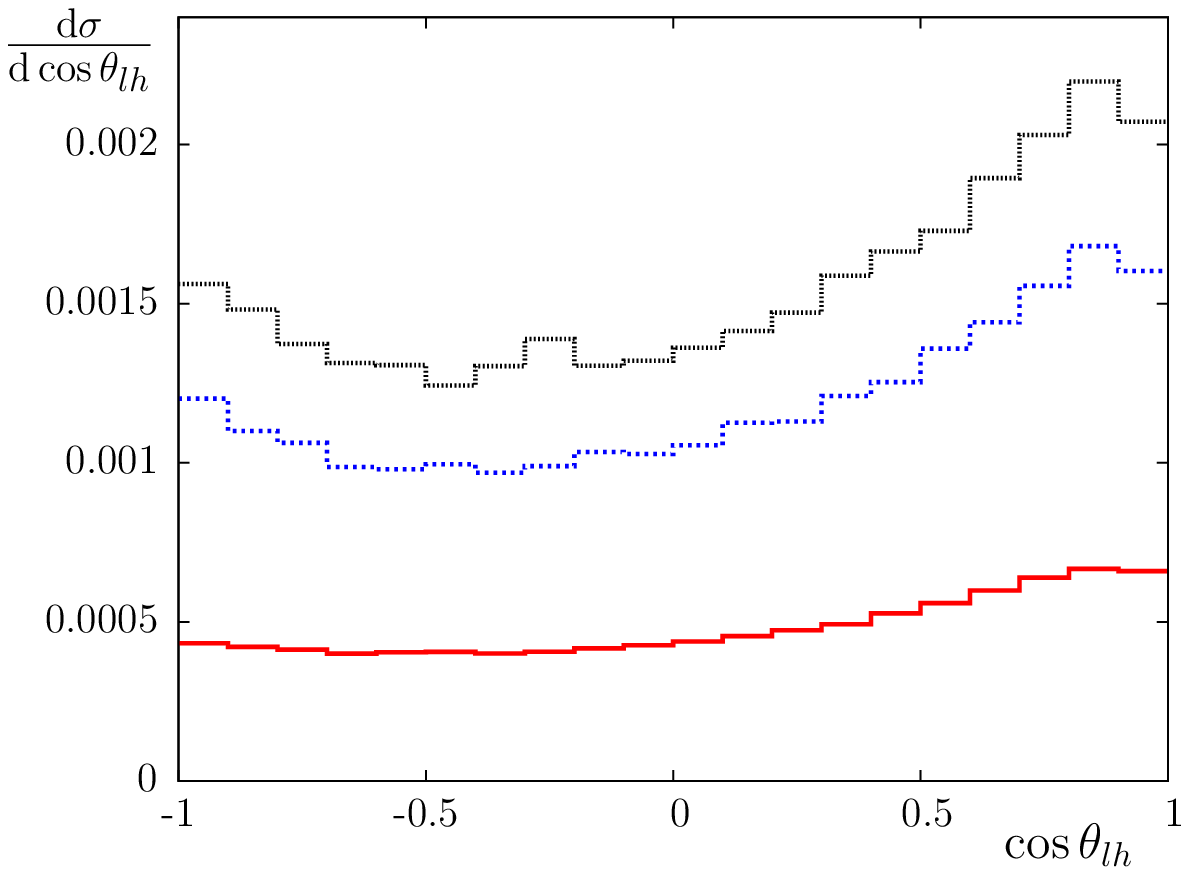}
\end{picture}
\end{center}
\vspace*{0.5cm}
\caption{The differential cross sections of (\ref{ppbbbudbmn}) at
$\sqrt{s}=14$~TeV as functions of the muon rapidity $y_l$, cosine
of the muon angle with respect to beam $\cos\theta_{l{\rm b}}$ and 
cosine of the muon angle with respect to the higgs boson $\cos\theta_{lh}$ 
computed with different cuts as described in the main text.}
\label{figsvb}
\end{figure}

\begin{figure}[htb]
\vspace*{1.5cm}
\begin{center}
\setlength{\unitlength}{1mm}
\begin{picture}(35,35)(0,0)
\includegraphics{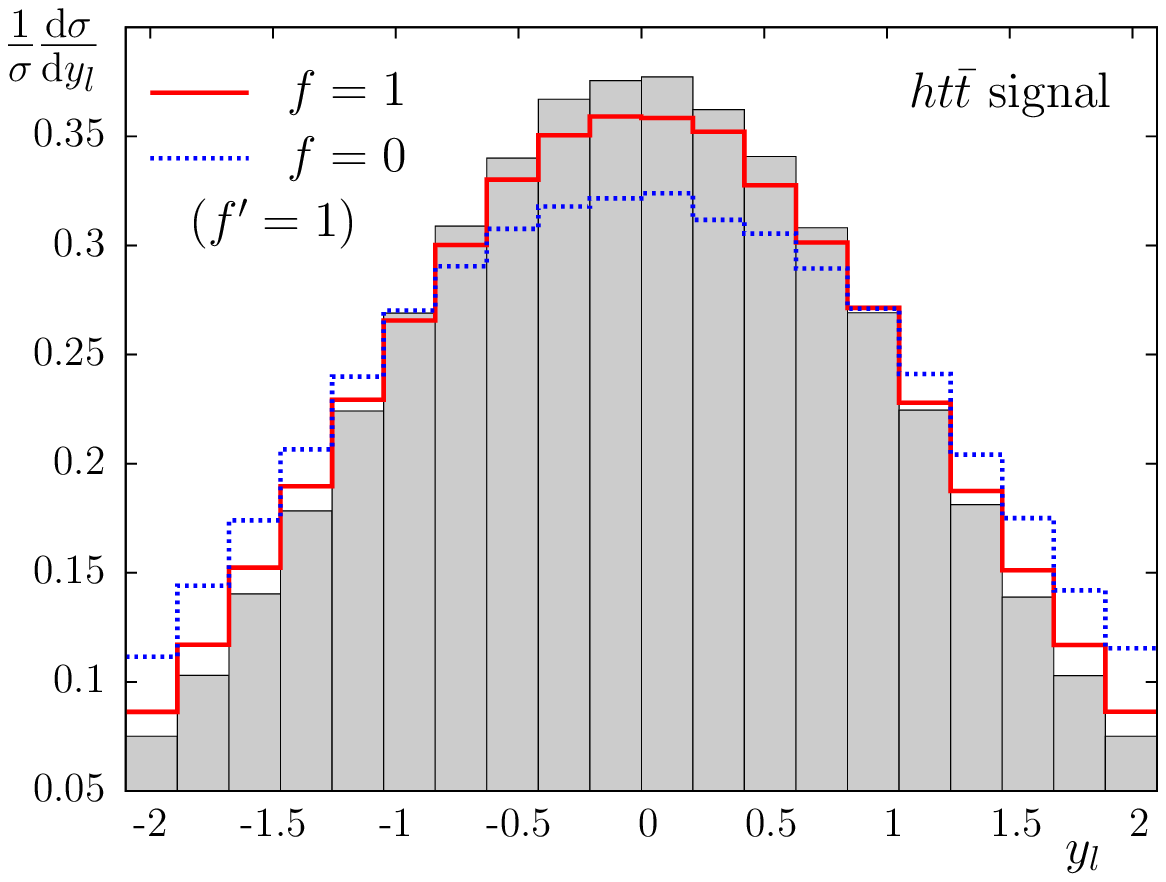}
\end{picture}
\hfill
\begin{picture}(35,35)(0,0)
\includegraphics{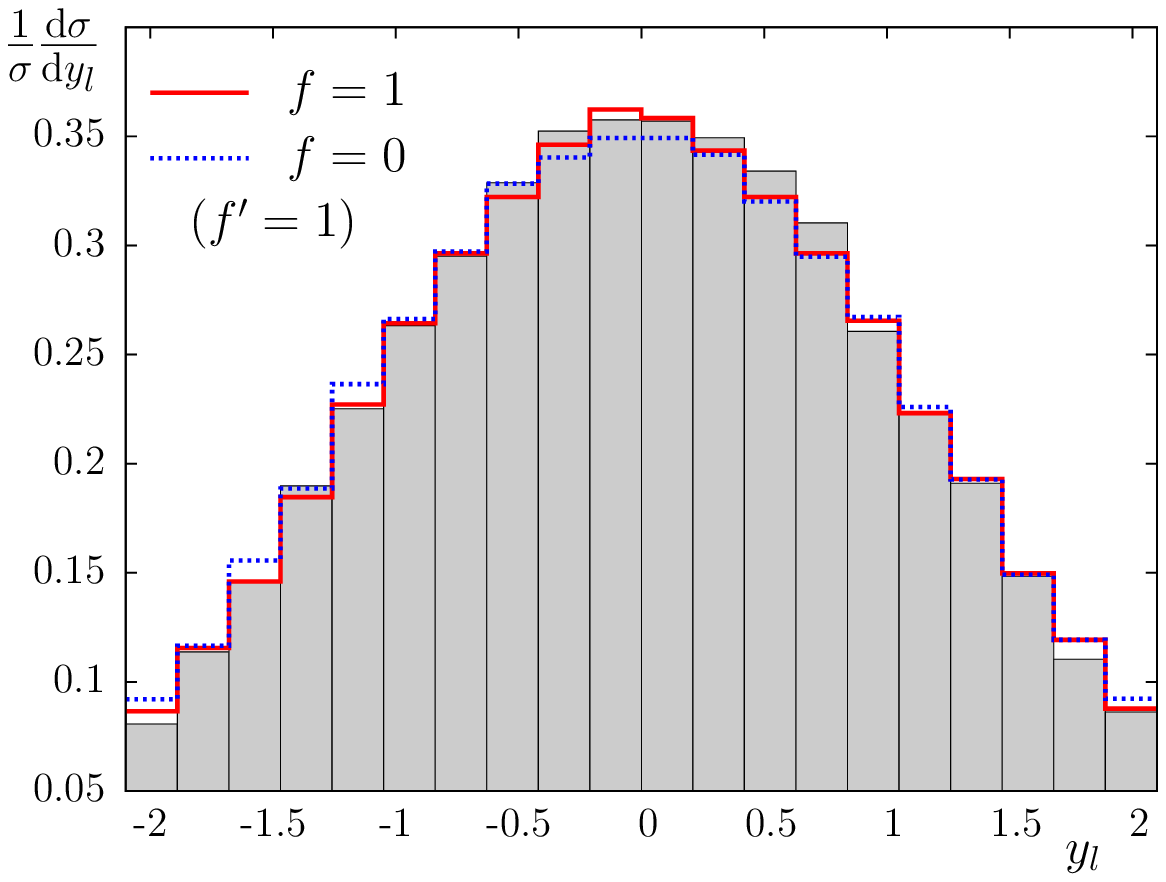}
\end{picture}\\[1.5cm]
\begin{picture}(35,35)(0,0)
\includegraphics{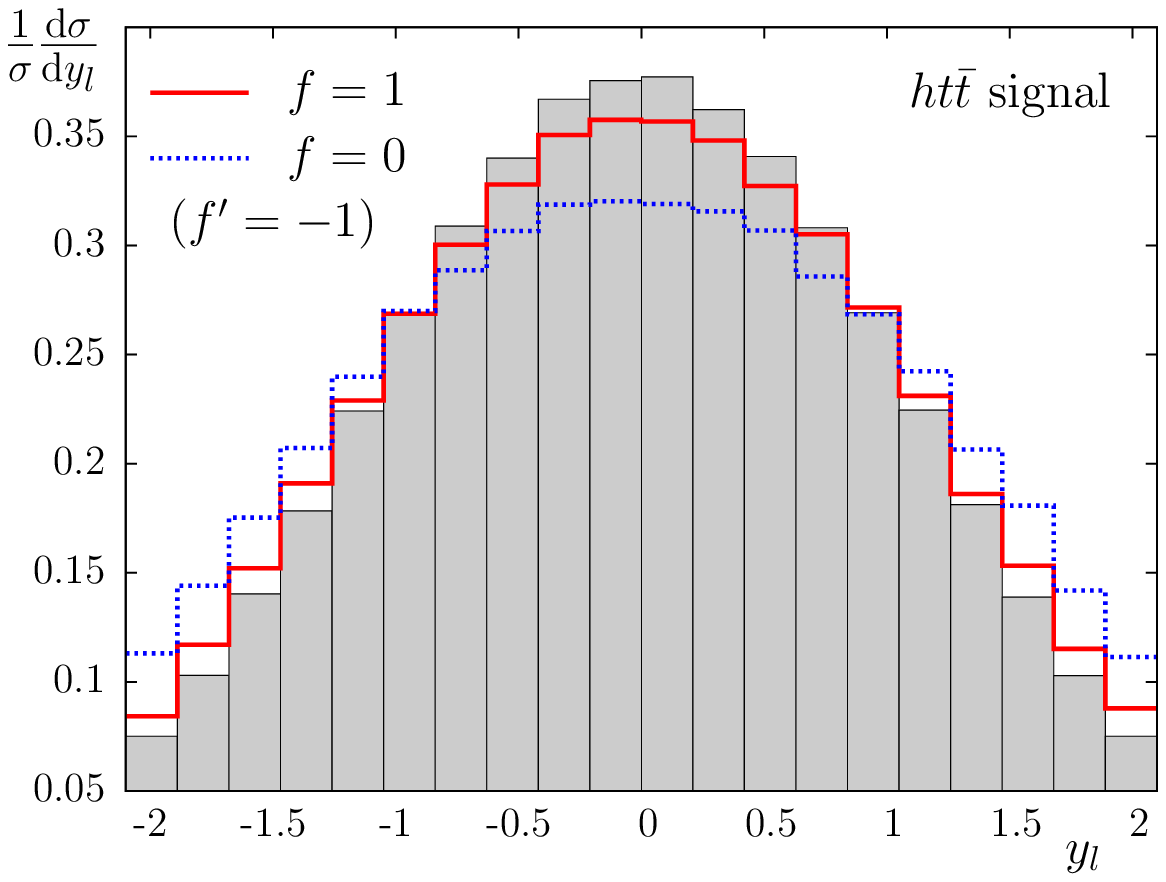}
\end{picture}
\hfill
\begin{picture}(35,35)(0,0)
\includegraphics{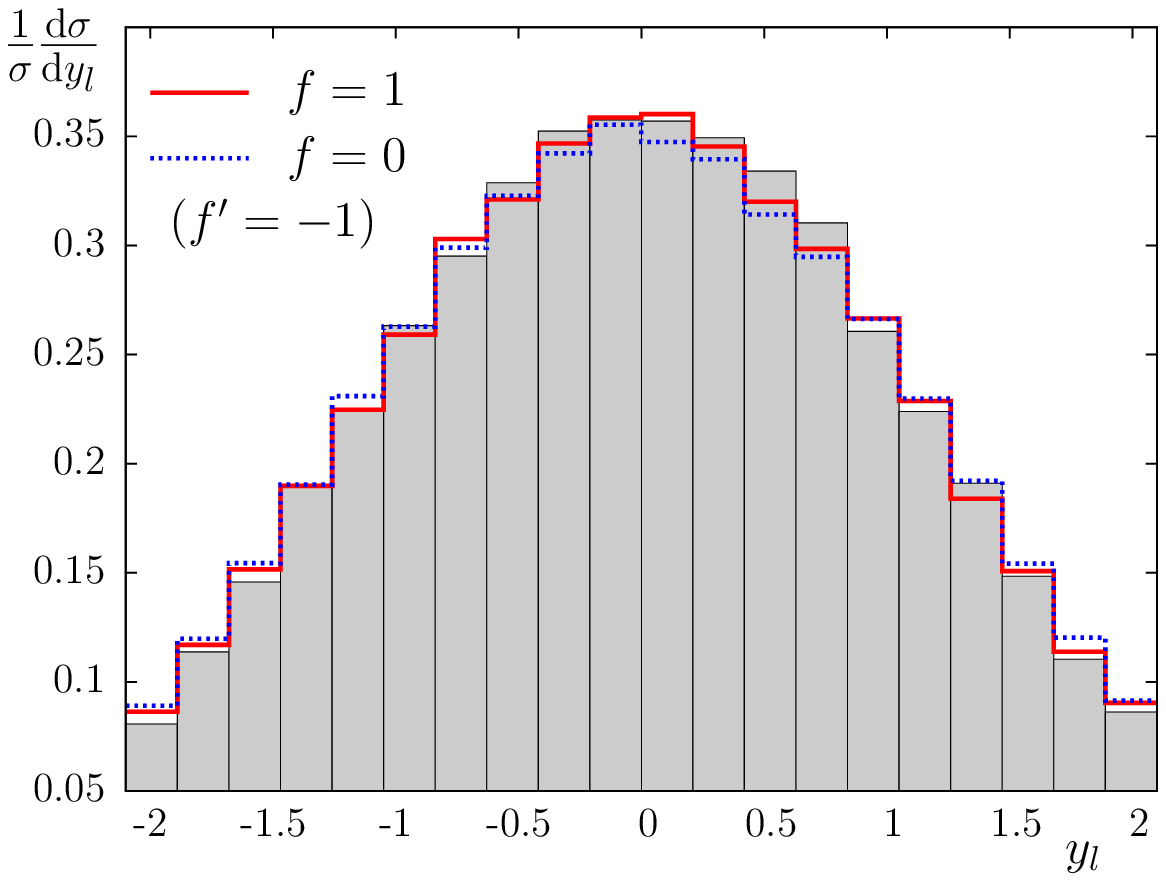}
\end{picture}
\end{center}
\vspace*{-1.cm}
\caption{Distributions in rapidity of the final state lepton of
(\ref{ppbbbudbmn}) in $pp$ collisions at $\sqrt{s}=14$~TeV with
different combinations of the scalar and pseudoscalar $t\bar th$
couplings: $t\bar th$ production signal (left panels) and
complete leading order prediction (right panels).
}
\label{figrapl}
\end{figure}

\begin{figure}[htb]
\vspace*{1.5cm}
\begin{center}
\setlength{\unitlength}{1mm}
\begin{picture}(35,35)(0,0)
\includegraphics{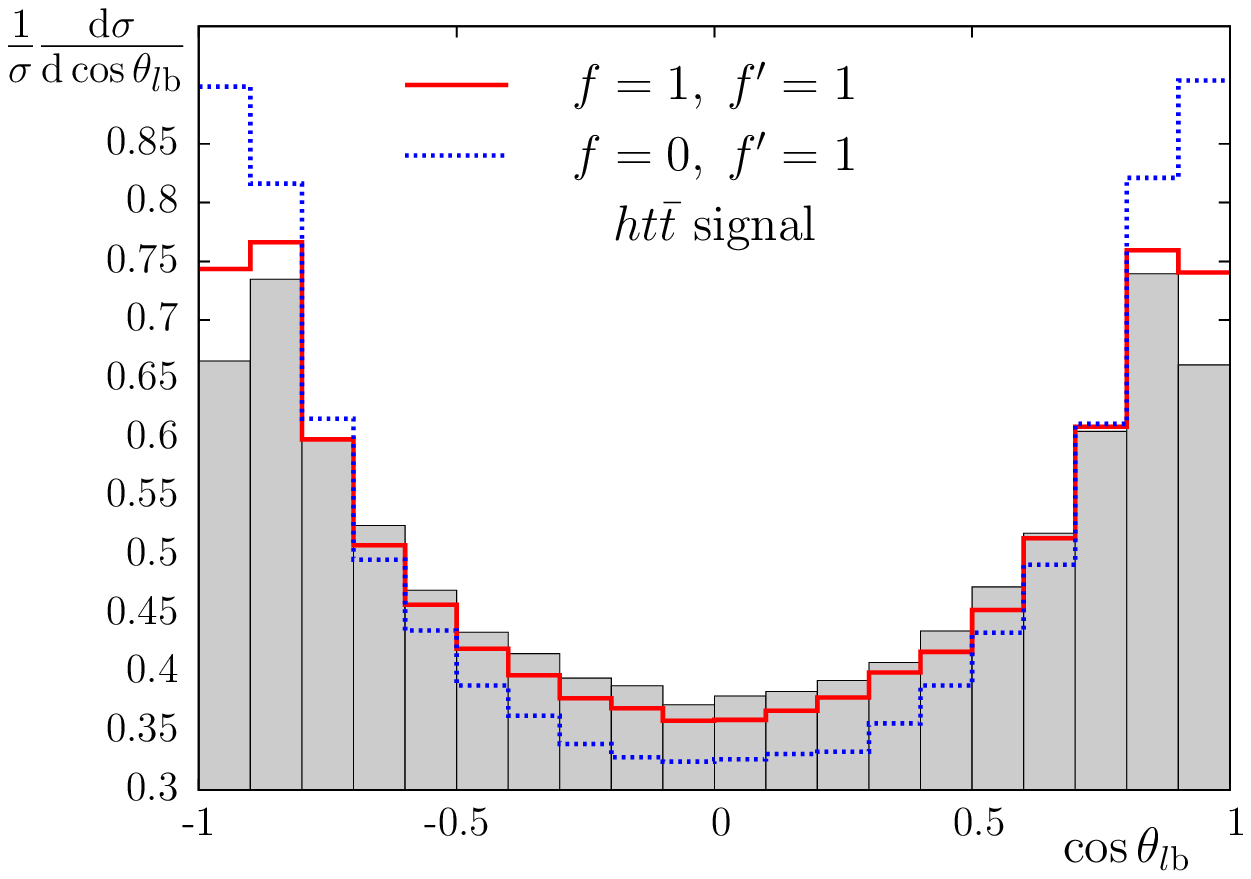}
\end{picture}
\hfill
\begin{picture}(35,35)(0,0)
\includegraphics{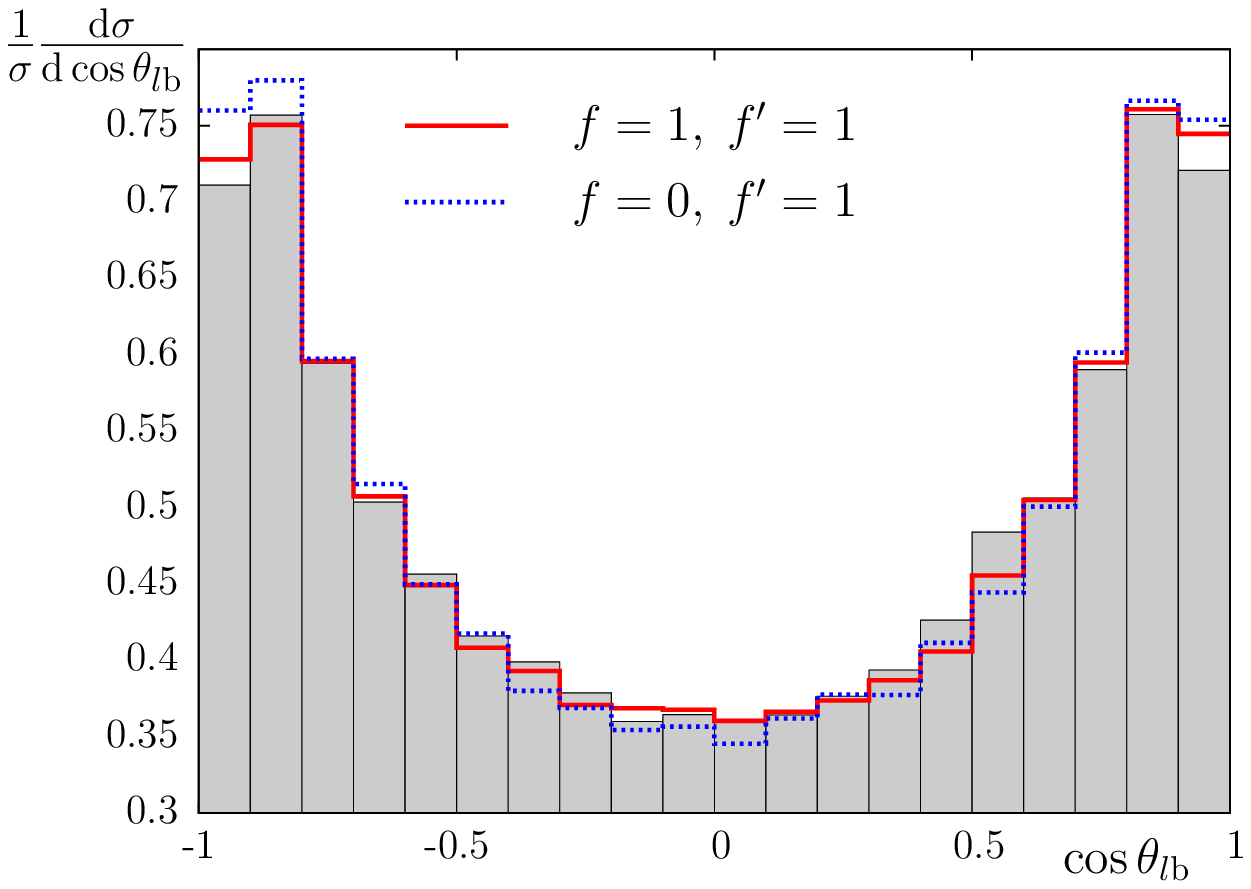}
\end{picture}\\[1.5cm]
\begin{picture}(35,35)(0,0)
\includegraphics{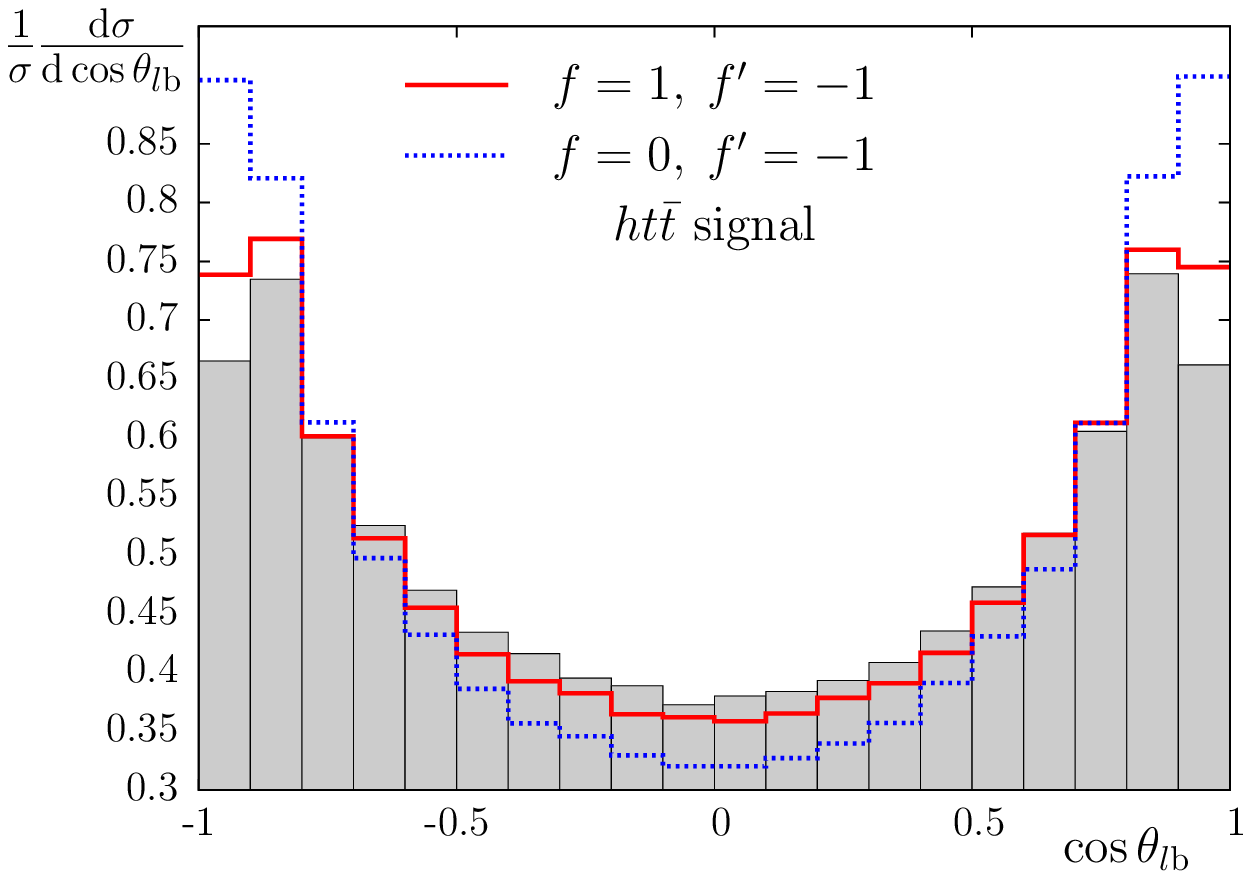}
\end{picture}
\hfill
\begin{picture}(35,35)(0,0)
\includegraphics{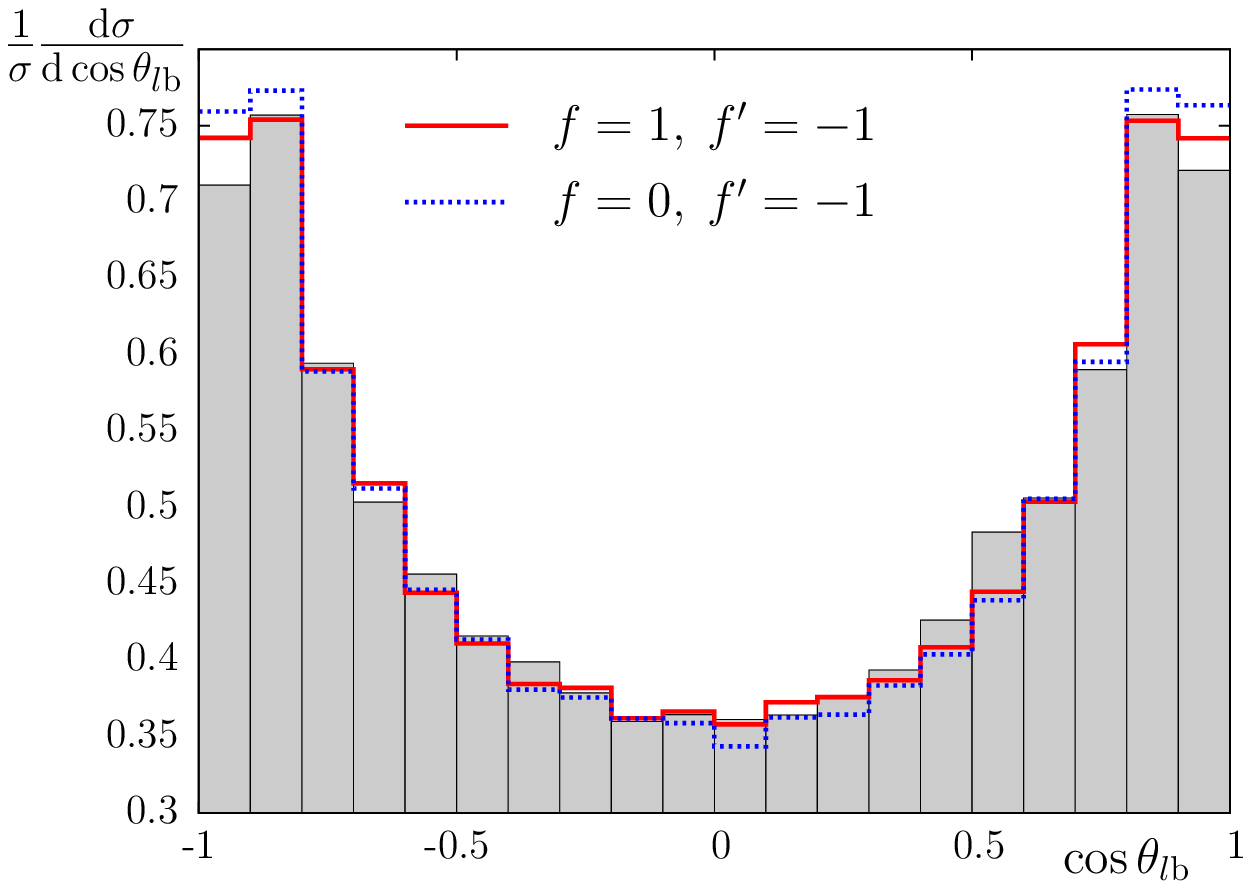}
\end{picture}
\end{center}
\vspace*{-1.cm}
\caption{Distributions in cosine of the angle between the final state lepton of
(\ref{ppbbbudbmn}) with respect to the beam in $pp$ collisions 
at $\sqrt{s}=14$~TeV with
different combinations of the scalar and pseudoscalar $t\bar th$
couplings: $t\bar th$ production signal (left panels) and
complete leading order prediction (right panels).
}
\label{figctlb}
\end{figure}

\begin{figure}[htb]
\vspace*{1.5cm}
\begin{center}
\setlength{\unitlength}{1mm}
\begin{picture}(35,35)(0,0)
\includegraphics{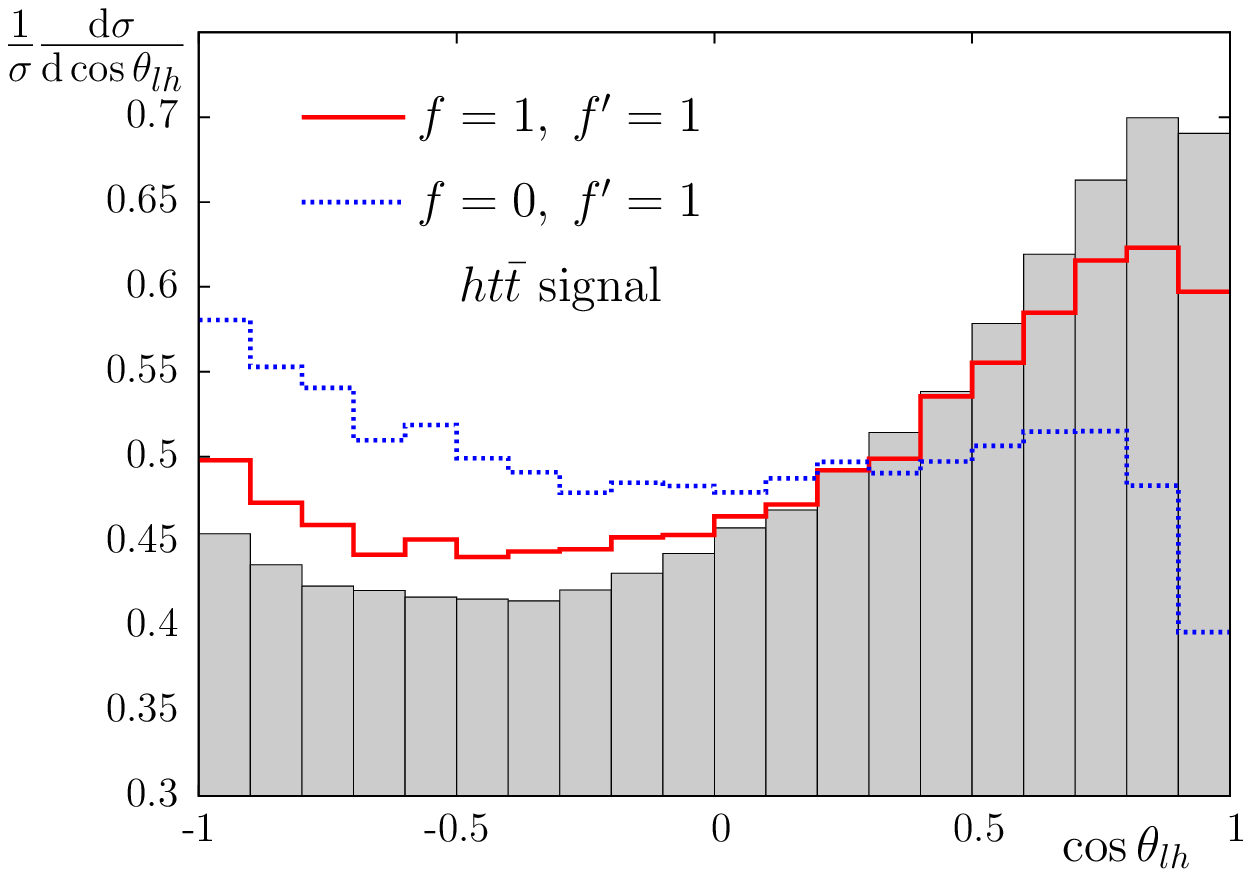}
\end{picture}
\hfill
\begin{picture}(35,35)(0,0)
\includegraphics{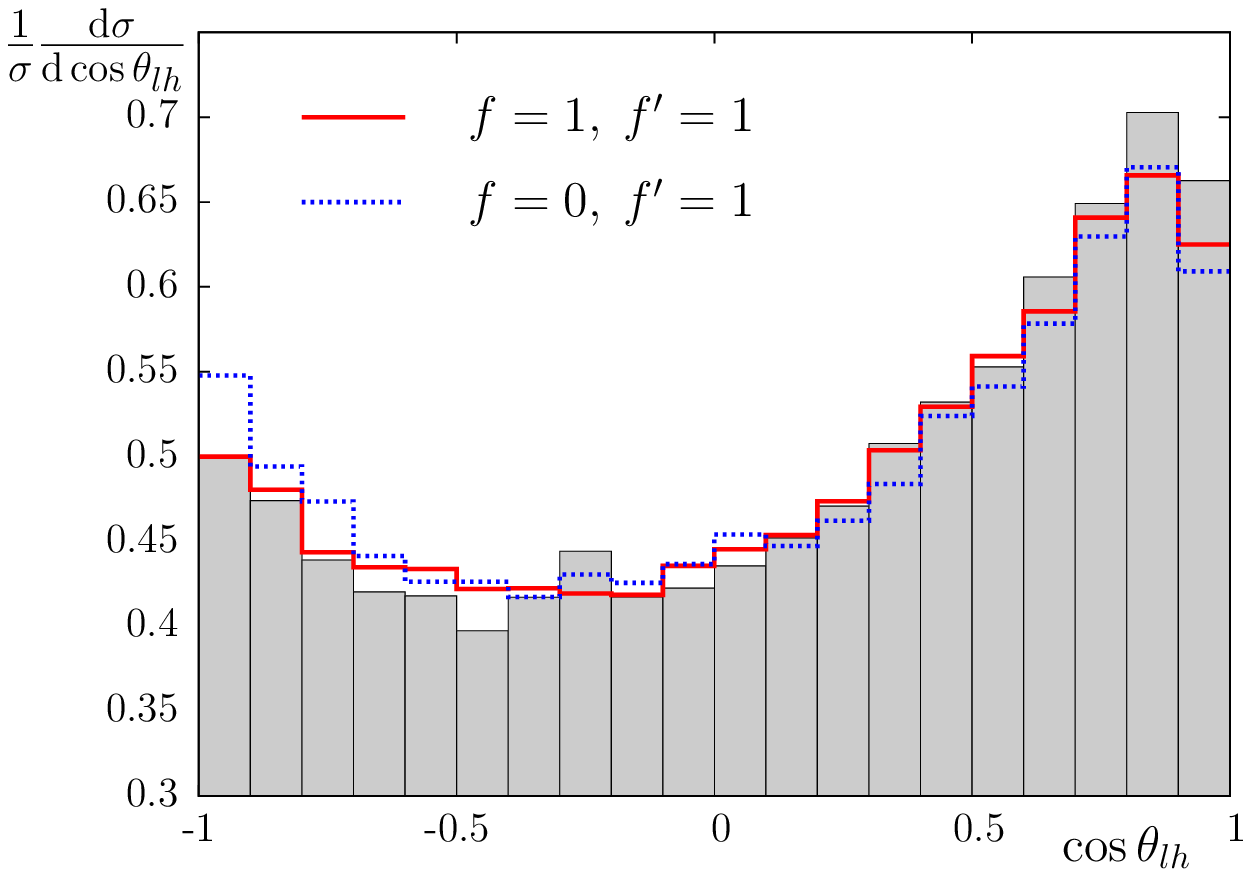}
\end{picture}\\[1.5cm]
\begin{picture}(35,35)(0,0)
\includegraphics{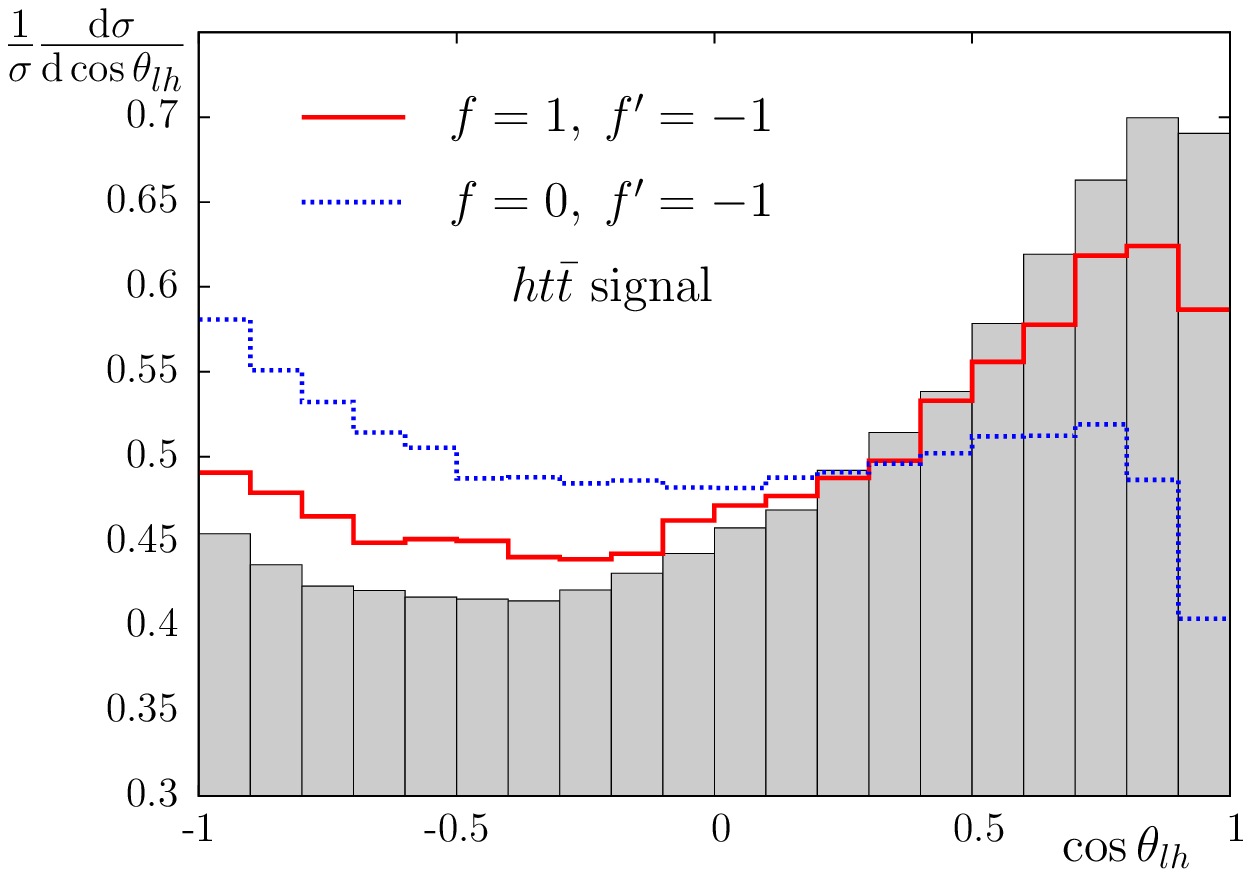}
\end{picture}
\hfill
\begin{picture}(35,35)(0,0)
\includegraphics{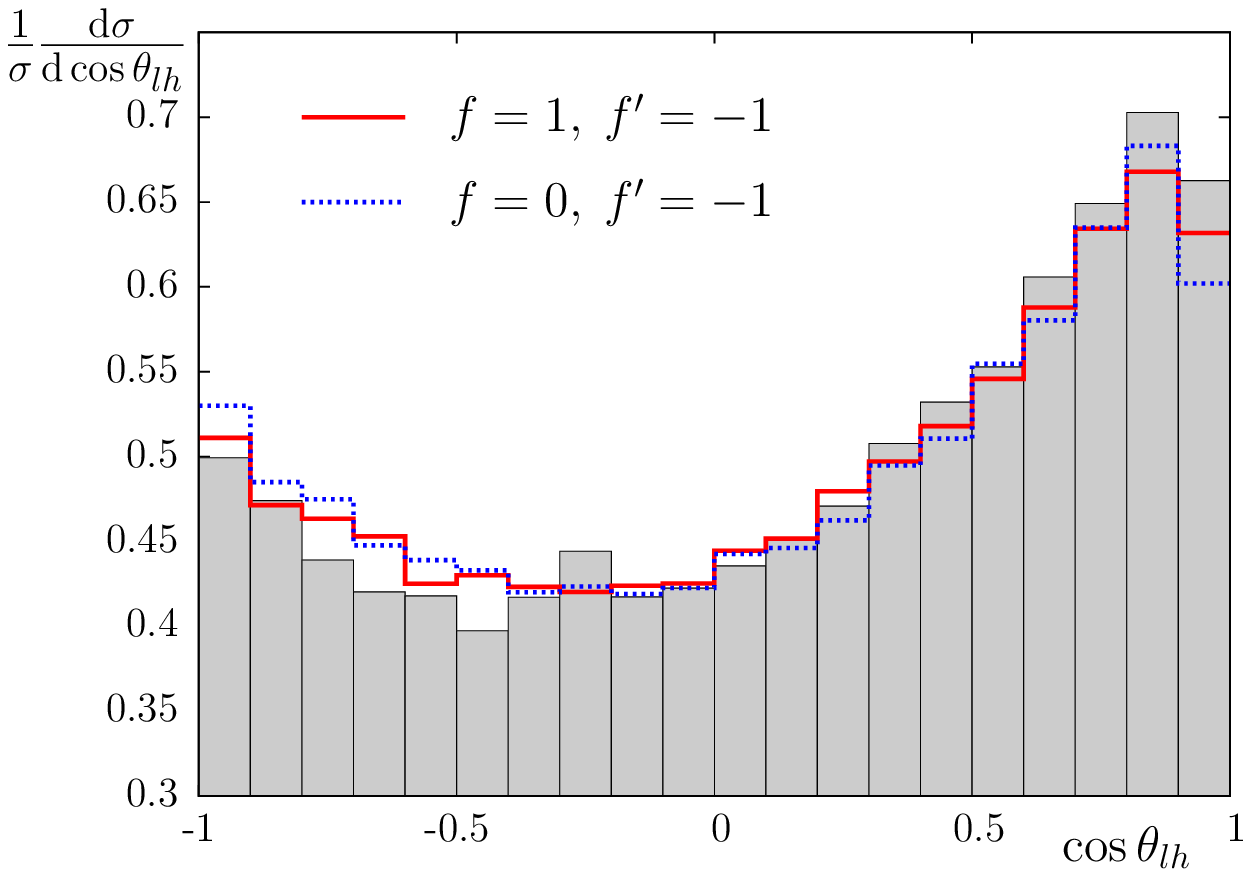}
\end{picture}
\end{center}
\vspace*{-1.cm}
\caption{Distributions in cosine of the angle between the final state lepton of
(\ref{ppbbbudbmn}) and the higgs boson in $pp$ collisions 
at $\sqrt{s}=14$~TeV with
different combinations of the scalar and pseudoscalar $t\bar th$
couplings: $t\bar th$ production signal (left panels) and
complete leading order prediction (right panels).
}
\label{figctlh}
\end{figure}

\begin{figure}[tb]
\vspace*{1.5cm}
\begin{center}
\setlength{\unitlength}{1mm}
\begin{picture}(35,35)(0,0)
\includegraphics{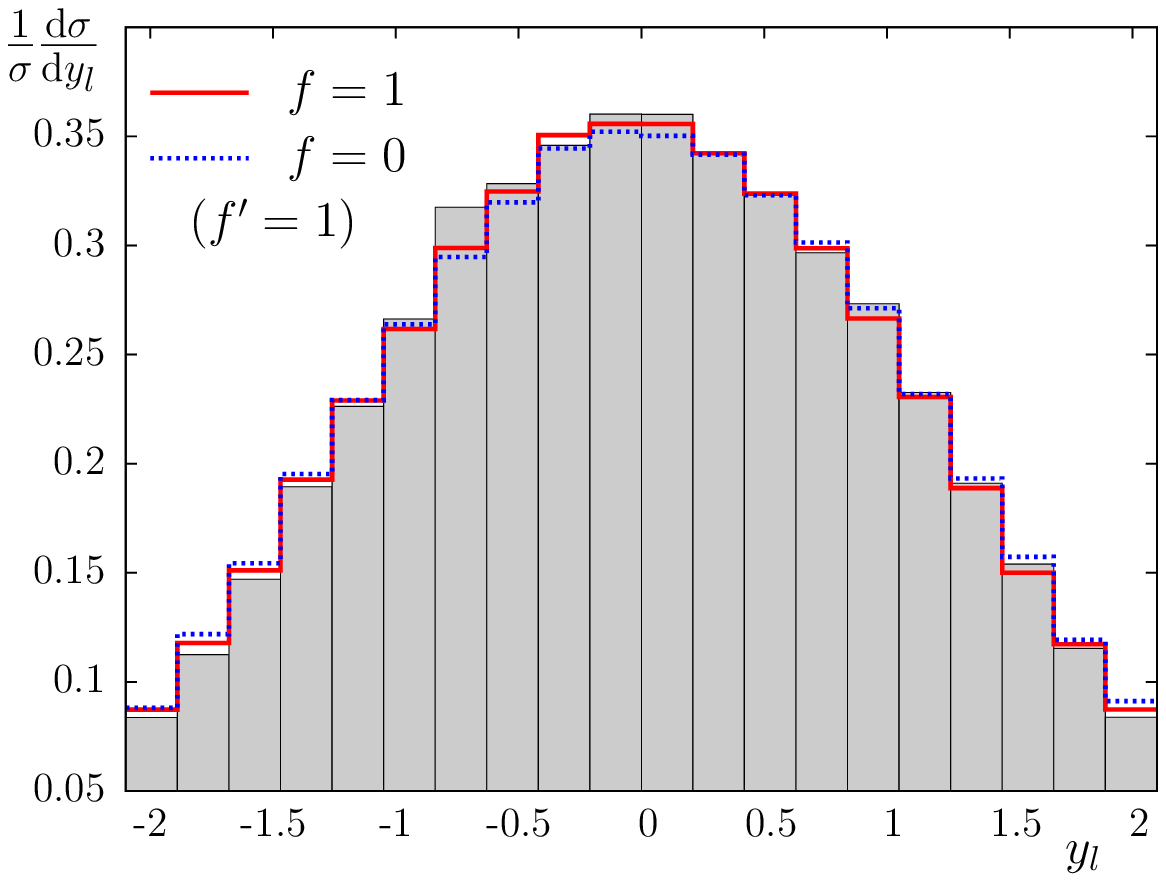}
\end{picture}
\hfill
\begin{picture}(35,35)(0,0)
\includegraphics{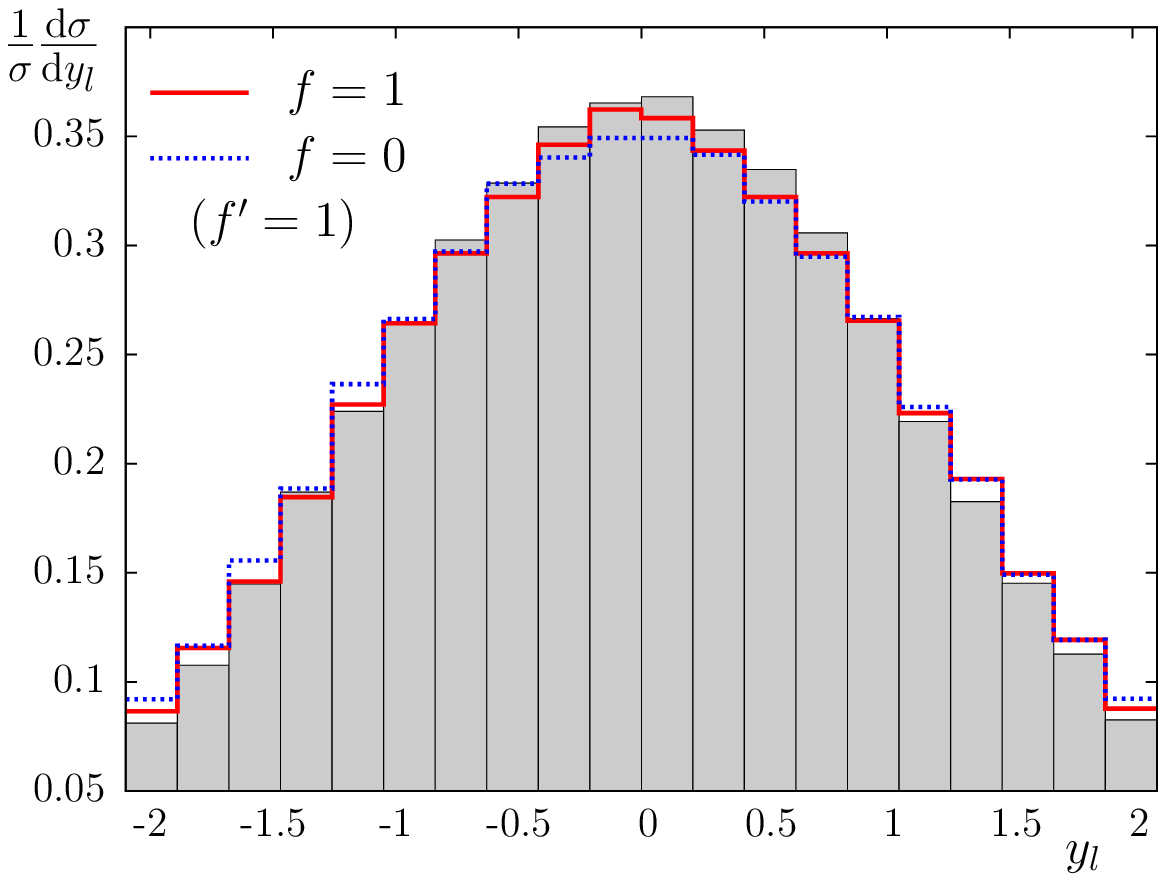}
\end{picture}\\[1.5cm]
\begin{picture}(35,35)(0,0)
\includegraphics{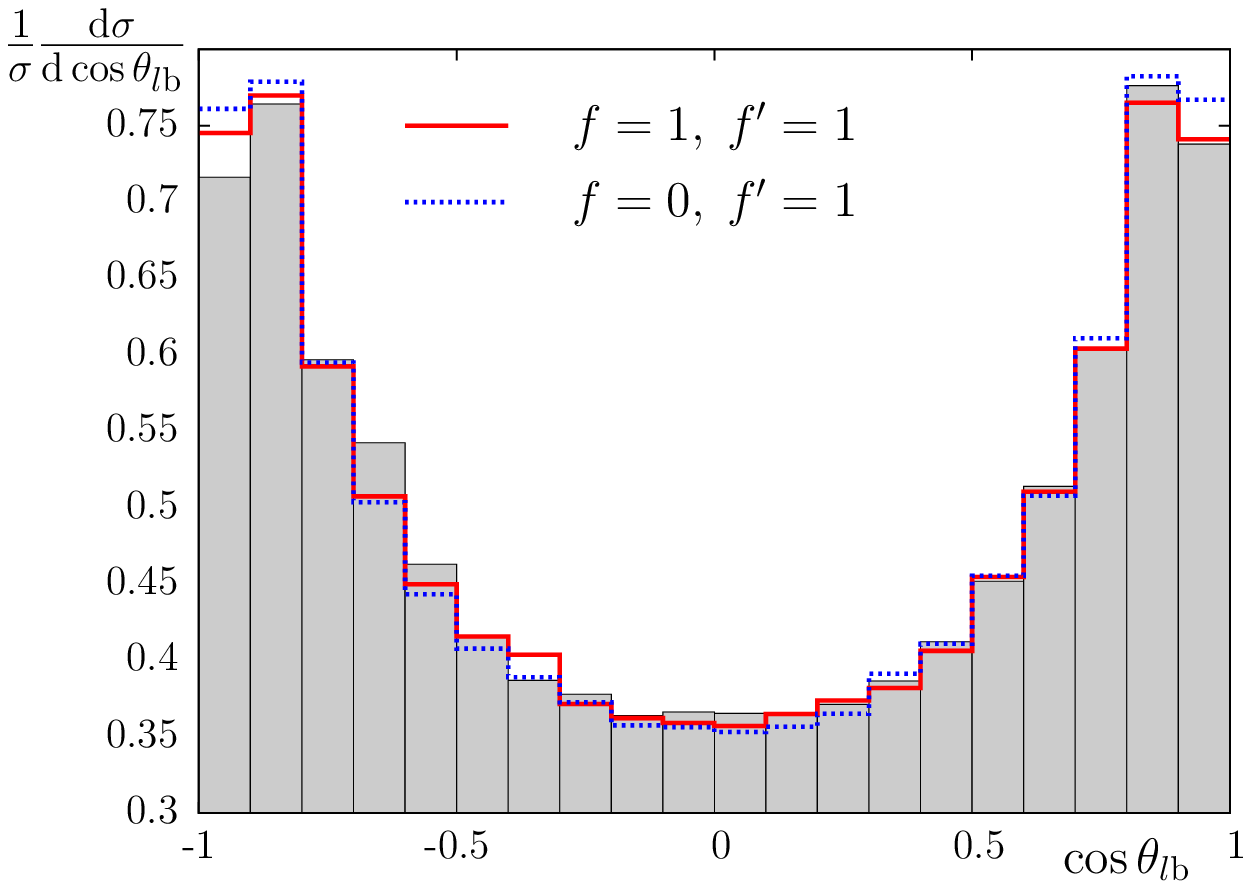}
\end{picture}
\hfill
\begin{picture}(35,35)(0,0)
\includegraphics{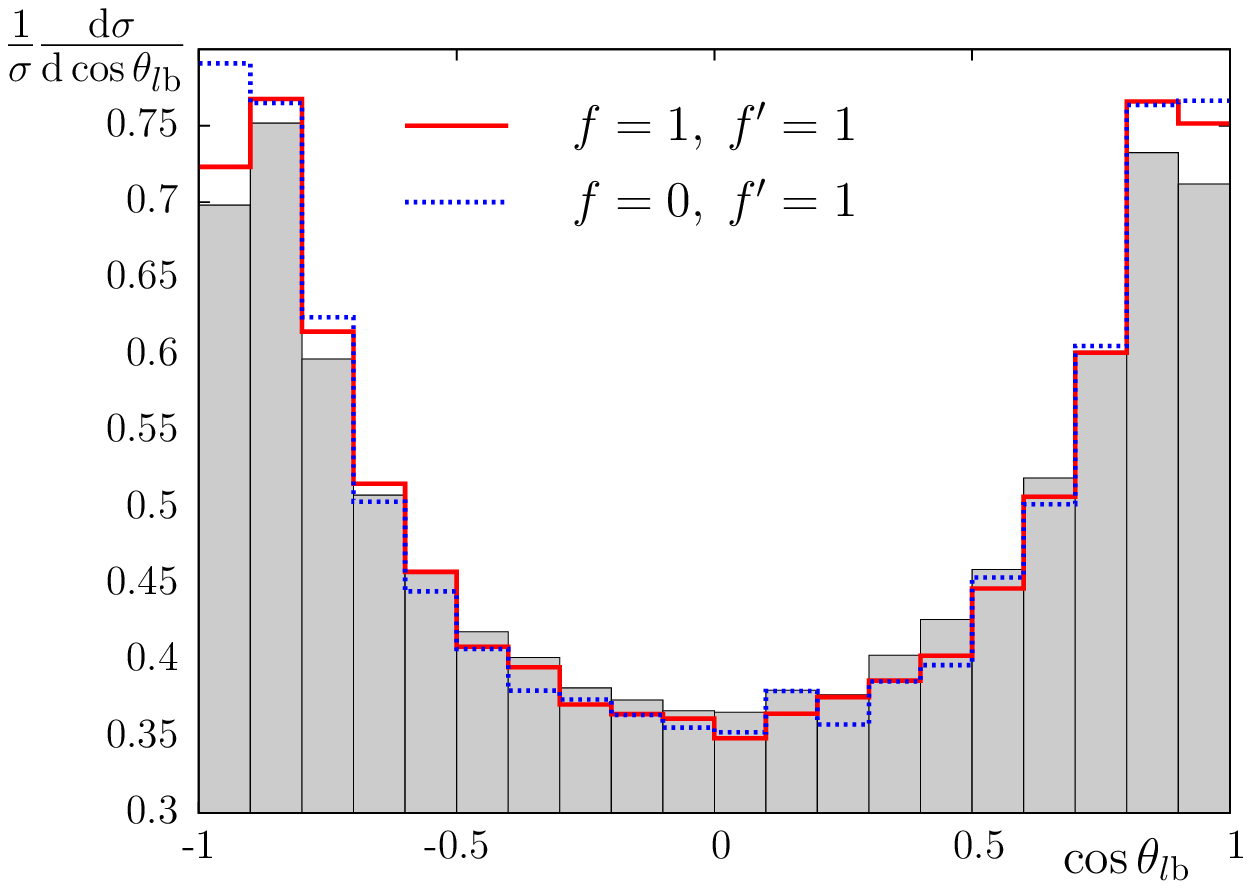}
\end{picture}
\end{center}
\vspace*{-1.cm}
\caption{Distributions of the final state lepton of (\ref{ppbbbudbmn}) 
at $\sqrt{s}=14$~TeV in $y_l$ and $\cos\theta_{l{\rm b}}$ with the cuts given by
(\ref{cuts}) (left panels) and (\ref{cuts})--(\ref{cutmtt}) and (\ref{cutmh}) 
with $m_{bb}^{\rm cut}=10$~GeV (right panels) for different combinations 
of the scalar and pseudoscalar $t\bar th$ couplings. 
}
\label{fig2d}
\end{figure}

The size of background contributions to the associated production of the 
higgs boson and top quark pair in $pp$ collisions at $\sqrt{s}=14$~TeV is 
illustrated in Fig.~\ref{figsvb}, where 
the differential cross sections of (\ref{ppbbbudbmn}) 
are plotted as functions of the muon 
rapidity, $y_l$, cosine of the muon angle with respect to beam,
$\cos\theta_{l{\rm b}}$ and cosine of the muon angle with respect to the 
reconstructed higgs boson momentum, $\cos\theta_{lh}$.
The cross sections plotted in Fig.~\ref{figsvb} have been computed with 
different cuts. In the left and central panel, the boxes shaded in light 
grey show the cross sections computed with cuts (\ref{cuts}) and
the grey boxes depict the cross sections calculated with cuts (\ref{cuts}) 
and the invariant mass cuts (\ref{cutmw})--(\ref{cutmtt}). These
results are not shown in the right panel of Fig.~\ref{figsvb}, 
as it is in principle not possible to reconstruct the higgs boson momentum 
without cut (\ref{cutmh}) on its invariant mass.
The short-dashed (dotted) lines show the results for $m_{bb}^{\rm cut}=20$~GeV 
($m_{bb}^{\rm cut}=10$~GeV) and the solid line in each panel of Fig~\ref{figsvb}
shows the results for the signal cross section calculated with
$m_{bb}^{\rm cut}=20$~GeV. It should be noted here that the signal cross
section is fairly independent of $m_{bb}^{\rm cut}$. 
As can be seen, the background is quite large and decreasing a value of 
$m_{bb}^{\rm cut}$ seems to be a right way towards its reduction.

Distributions in 3 different kinematical variables of the final state lepton
of (\ref{ppbbbudbmn}): rapidity, $y_l$,
cosine of the angle with respect to the beam, $\cos\theta_{l{\rm b}}$ and
cosine of the angle with respect to the reconstructed higgs boson momentum,
$\cos\theta_{lh}$ are shown in Figs.~\ref{figrapl}--\ref{figctlh}.
In each of the figures, the left panels show the $t\bar th$ production signal, 
computed with 56 Feynman diagrams, and the
right panels show the complete leading order predictions, computed
with $67\,300$ Feynman diagrams. The grey boxes show the corresponding 
SM results, i.e. the results obtained with $f=1$ and $f'=0$.
Cuts (\ref{cuts})--(\ref{cutmtt}) and (\ref{cutmh}) with
$m_{bb}^{\rm cut}=20$~GeV are applied to all the distributions 
presented.
A relatively clear effect of the anomalous couplings $f$ and $f'$ 
that can be seen in the $t\bar th$ signal distributions on the left hand side
of all Figs.~\ref{figrapl}--\ref{fig2d} is to large extent
obscured by the background contributions in the plots on the right hand side
which show the full leading order results. 

In order to illustrate what a role invariant mass cuts 
(\ref{cuts})--(\ref{cutmh}) play,
the distributions of the final state lepton of (\ref{ppbbbudbmn}) 
at $\sqrt{s}=14$~TeV in $y_l$ and $\cos\theta_{l{\rm b}}$ with different
cuts are compared in Fig.~\ref{fig2d}. The left panels show
the distributions with the cuts given by (\ref{cuts}) and the right
panels show the distributions with cuts (\ref{cuts})--(\ref{cutmtt}) 
and (\ref{cutmh}) with $m_{bb}^{\rm cut}=10$~GeV for two different 
combinations of the scalar and pseudoscalar $t\bar th$ couplings of 
(\ref{httcoupl}). Again the grey boxes show the SM results. Although the 
invariant mass cuts suppress the
background contributions to some extent, the degree of suppression does
not look very satisfactory.

The signal significance $\mu=\sigma(f,f')/\sigma_{SM}$ and corresponding 
differences in expected numbers of events, $\Delta n=n(f,f')-n_{SM}$, 
for different 
combinations of the couplings $(f,f')$, with the $\mu^-$ in the
forward $(\cos\theta_{lh}>0)$ or backward $(\cos\theta_{lh}<0)$ hemisphere
with respect to the direction of the higgs boson,
in reaction (\ref{ppbbbudbmn}) at $\sqrt{s}=14$~TeV are shown in 
Table~\ref{Tab}. The cuts are given by (\ref{cuts})--(\ref{cutmh}) 
with $m_{bb}^{\rm cut}=20$~GeV (columns 2--5) and 
$m_{bb}^{\rm cut}=10$~GeV (columns 6--9). The event numbers have been 
calculated assuming
an integrated luminosity of $100~{\rm fb}^{-1}$ and 100\% detection efficiency.
Therefore, they should be treated with care, in particular because of
the fact that only the leading order contributions to 
reaction (\ref{ppbbbudbmn}) are taken into account. However, the leading order
predictions for signal significance $\mu$ are more reliable. 
In particular, $\mu\approx 1.2$ for $f=1$ and $|f'|=1$ indicates 
a potential of the reaction of associated production of the higgs boson and 
top quark pair in obtaining direct limits on the 
pseudoscalar coupling $f'$. If only the $t\bar t h$ signal contributions to
the cross section are taken into account, then the signal significance 
for this particular combination of couplings becomes 
even bigger, amounting to $\mu = 1.4$ in the forward and
$\mu= 1.6$ in the backward hemisphere with respect to the direction
of the higgs boson. This again shows how the off resonance background
contributions obscure the signal of $t\bar t h$ production.

\begin{table}[!ht]
\begin{center}
\begin{tabular}{c|rrrr|rrrr}
\hline 
\hline 
\rule{0mm}{6mm}
&\multicolumn{4}{c|}{$m_{bb}^{\rm cut}=20$~GeV} 
&\multicolumn{4}{c}{$m_{bb}^{\rm cut}=10$~GeV}\\ [1.5mm]
& \multicolumn{2}{|c}{$\cos\theta_{lh}<0$} & 
  \multicolumn{2}{c|}{$\cos\theta_{lh}>0$} & 
\multicolumn{2}{c}{$\cos\theta_{lh}<0$} & 
  \multicolumn{2}{c}{$\cos\theta_{lh}>0$} \\[1.5mm] 
\cline{2-9}
\rule{0mm}{6mm}
$(f,f')$ 
&$\mu\;\;$&$\Delta n$&$\mu\;\;$&$\Delta n$&$\mu\;\;$&$\Delta n$&$\mu\;\;$&$\Delta n$\\[1.5mm] 
\hline
\rule{0mm}{6mm} $(0,1)$ &0.90 &$-148$ &0.83 &$-265$ &0.85 &$-174$ 
&0.78 &$-275$
\\[1.5mm] 
\rule{0mm}{6mm} $(0,-1)$ &0.90 &$-151$ &0.84 &$-252$ &0.84 &$-188$ &0.78 &$-270$ \\[1.5mm] 
\rule{0mm}{6mm} $(1,1)$ &1.20 &295 &1.17 &251 &1.23 &261 &1.17 & 210 \\[1.5mm] 
\rule{0mm}{6mm} $(1,-1)$ &1.20 &302 &1.15 &238 &1.24 &279 &1.18 & 221
\end{tabular} 
\end{center}
\caption{The signal significance $\mu$ and corresponding difference in numbers 
of events $\Delta n$ for different combinations of the couplings 
$(f,f')$ in reaction (\ref{ppbbbudbmn}) at $\sqrt{s}=14$~TeV.}
\label{Tab}
\end{table}

\section{Summary and conclusions}

The differential cross sections and distributions of the final state lepton 
of (\ref{ppbbbudbmn}) in rapidity, 
cosine of its angle with respect to the beam and
cosine of its angle with respect to the reconstructed higgs boson momentum
have been computed
to the leading order in the presence of most general $t\bar th$ interaction 
with operators of dimension-six.
The distributions computed with the $t\bar th$ signal
diagrams only, which are substantially changed in the presence of 
anomalous $t\bar th$ couplings, 
have been compared with those computed with the full set
of the leading order Feynman diagrams. The comparison
have shown that the background contributions to large extent
obscure the relatively clear effects of the
anomalous $t\bar th$ coupling in the signal distributions. 
This means that analyses of such effects \cite{degrande}, in addition to 
higher order corrections \cite{czakon} that are usually 
calculated for the on-shell top quarks and higgs boson, 
should include their decays and possibly complete off resonance background 
contributions to the corresponding exclusive reactions.
The only reasonable way to make the effects of anomalous couplings better 
visible seems to be imposing more and more restrictive cuts.

Acknowledgements: This project was supported in part with financial resources 
of the Polish National Science Centre (NCN) under grant decision 
number DEC-2011/03/B/ST6/01615 and by the Research Executive 
Agency (REA) of the European Union under the Grant Agreement number 
PITN-GA-2010-264564 (LHCPhenoNet).

\end{document}